\documentclass[runningheads]{llncs}  

\usepackage{graphicx}
\usepackage{amsmath}
\usepackage{algorithm}
\usepackage{algorithmic}
\usepackage{url}
\urldef{\mailsa}\path|{chowdary, psk}@cse.iitm.ac.in|

\begin{document}

\mainmatter  % start of an individual contribution

\title{USUM: Update Summary Generation System}

\titlerunning{USUM: Update Summary Generation System}


\author{C Ravindranath Chowdary \and P Sreenivasa Kumar}
\authorrunning{C Ravindranath Chowdary and P Sreenivasa Kumar}
%(feature abused for this document to repeat the title also on left hand pages)

% the affiliations are given next
\institute{Department of Computer Science and Engineering\\ Indian Institute of Technology Madras \\Chennai 600 036, India\\
\mailsa\\
\url{}}


\toctitle{USUM: An Efficient Update Summary Generation System}
\tocauthor{C Ravindranath Chowdary and P Sreenivasa Kumar}
\maketitle

\begin{abstract}
\emph{
Huge amount of information is present in the World Wide Web  and a large amount is being added to it frequently. A query-specific summary of multiple documents is very helpful to the user in this context. Currently, few systems have been proposed for query-specific, extractive multi-document summarization. If a summary is available for a set of documents on a given query and if a new document is added to the corpus, generating an updated summary from the scratch is time consuming and many a times it is not practical/possible. In this paper we propose a solution to this problem. This is especially useful in a scenario where the source documents are not accessible. We cleverly embed the sentences of the current summary into the new document and then perform query-specific summary generation on that document. Our experimental results show that the performance of the proposed approach is good in terms of both quality and efficiency.
}
\end{abstract}
\section{Introduction}
\label{introduction}
Currently, the World Wide Web is the largest source of information. Huge amount of data is present on the Web and large amount of data is added to the web constantly. Often the information pertaining to a topic is present across several web pages. It is a tedious task for the user to go through all these documents as the number of documents available on a topic will range from tens to thousands. It will be of great help for the user if a query specific multi-document summary is generated. Summary generation can be broadly divided as abstractive and extractive. In abstractive summary generation, the abstract of the document is generated. The summary so formed need not have exact sentences as present in the document. In extractive summary generation, important sentences are extracted from the document. The generated summary contains all such extracted sentences arranged in a meaningful order. In this paper, generated summaries are extractive. Summary can be generated either on a single document or on several documents. In multi-document summary generation, other issues like time, ordering of extracted sentences, scalability etc. will arise. 

Summary can be either generic or query specific. In generic summary generation, the important sentences from the document are extracted and the sentences so extracted are arranged in appropriate order. In query specific summary generation, the sentences are scored based on the query given by the user. The highest scored sentences are extracted and presented to the user as summary. For a set of documents on a topic and a query related to the topic, suppose a summary is available. If a new document is now made available to the system then summary has to be regenerated with the new document included into the input set by running the query-specific summarizer. But this is not a good solution as it takes considerable amount of time to run the summarizer afresh and a lot of space to store all the original documents. Also, most of the times the documents used for summarization may not be accessible. 

Broadly the methodology used to summarize multiple documents is to combine all the documents into a unified structure in an intelligent (each system/approach has its own methodology) fashion and then the summary is generated by taking the unified structure as the input. The construction of unified structure is a time taking task. So, the time complexity of query-specific multi-document summarization is high. 

In this paper we address the following problem: given an extractive summary that is generated for a given query on a set of documents, upon the arrival of a new document, the summary has to be updated without considering the initial set of documents. The proposed system will be given the present summary and the ($(n+1)^{th}$) document as the input and the output should be the updated summary. In this paper we propose a novel and efficient model for query-specific update summary generation using extractive mechanism. To the best of our knowledge this problem is not addressed in the literature.     

The rest of the paper is organized as follows: In Section \ref{RelatedWork} we discuss the related work. Generation of embedded document is discussed in Section \ref{Model}. In Section \ref{Algorithm} we introduce the methodology to accomplish the task of update summary generation. Experimental setup is discussed in Section \ref{ExperimentalSetup}. In Section \ref{ExperimentalResults} results are discussed and conclusions are given in Section \ref{Conclusions}.

\section{Related Work}
\label{RelatedWork}
Text summarization has gained popularity in the recent years. 
A generic summary generation on single document is discussed by Cajun Wan et al. \cite{Iterative_Reinforcement_Approach}. Both summary and keywords are extracted from a single document by following iterative reinforcement approach. To extract summary from the document, the following relations are used: sentence-sentence relation, word-word relation and sentence-word relation. A generic summary generation on multiple documents is discussed by Radev et al. in  \cite{MEAD}. Centroid based approach is followed by this system, called MEAD, to generate summary. Given a set of documents about a particular topic i.e., a cluster of documents, the centroid of the cluster is calculated. A score is given to each sentence in the cluster with respect to the centroid. Sentences are selected in decreasing order of sentence scores and are arranged with respect to the chronological order of their respective documents. 

Extractive summary generation is discussed in \cite{MEAD,ML_ClASSY,QueSTS,Varadarajan}. The input to extractive summarizers is the set of documents that are to be summarized and the output is the sentences extracted from the input documents. The sentences so extracted are arranged in a manner which increases coherence (logical flow) to the generated summary. In particular, the former criteria is addressed in \cite{QueSTS}.  

Single document generic summary is discussed in \cite{Iterative_Reinforcement_Approach}, here, extraction of the sentences from a document to generate a summary is accomplished by using sentence-sentence, word-word and sentence-word relationships. Single document query-specific summary generation is discussed in \cite{Varadarajan}, here, a connected sub-graph of sentences are extracted from the document graph. Sentences are said to be connected if the similarity measure between them is above a threshold. Multi-document generic summary generation is discussed in \cite{MEAD,LexPageRank}. In \cite{MEAD}, all the sentences from the documents are given scores and the sentences are selected into the summary in the decreasing order of their scores. In \cite{LexPageRank}, sentences are given scores based on the model inspired by PageRank \cite{PageRank}.    

Multi-document query-specific summary generation is discussed in \cite{ManifoldRanking,QueSTS}. In \cite{ManifoldRanking}, query is also considered as one of the sentences in a document. Similarities between all the pairs of sentences in the documents are calculated and these similarity values are used while giving the scores to the individual sentences. In \cite{QueSTS}, two types of scores are calculated, first one is based on the similarity between sentences and the second is based on the similarity of a sentence with respect to the query. 

Centrality based approaches are discussed in \cite{Salton,LexPageRank,TextRank,Rada_GraphBased}. In centrality based approaches, the salience of a sentence is calculated based on both the contribution of the sentence and the type of neighbouring sentences it is surrounded. Degree centrality is discussed in \cite{Salton} and eigenvector centrality is discussed in  \cite{LexPageRank,TextRank,Rada_GraphBased}. Concept of bushy path was introduced by Salton et al. in \cite{Salton}. Nodes with high degree are called bushy nodes. Bushy path is defined as a path connecting top $n$ bushy nodes. Eigenvector centrality of a node is calculated by taking into consideration both the degree of the node and the degree of the nodes connecting to it. This is inspired by PageRank \cite{PageRank}.

Redundancy handling is addressed in \cite{MMR}. This principal is followed by many other systems. Mean marginal relevancy(MMR) principal is as follows: Node scores are calculated w.r.t the query. Summary is generated incrementally. A node with highest score is selected into the summary. All the scores of remaining nodes  are recalculated based on the nodes already selected into summary and the node score they possess. From the recalculated scores, the highest scored node will be added to summary.  

In all the above approaches, a summary is generated from scratch. In this paper we address the problem of updating the extracted summary with the availability of a new document. Here we update the summary for a given query. This problem of update summary generation is proposed by us and the  detailed procedure to accomplish this task is explained in the following sections.
   
\section{Generating Summary-Embedded Document}
\label{Model}
%embidding summary into doc
We follow a graph based approach to accomplish the task of update summary generation. Every sentence in the document is a node and the edges are placed between the nodes if the similarity score between them is above a threshold. Hereafter we use the words, ``node" and ``sentence", interchangeably. Similarity between the nodes is calculated using the Equation \ref{EdgeScore}.
\begin{equation}
\centering
\label{EdgeScore}
sim(\overrightarrow{n_i},\overrightarrow{n_j}) = \frac{\overrightarrow{n_i}.\overrightarrow{n_j}}{|\overrightarrow{n_i}|  |\overrightarrow{n_j}|}
\end{equation} 
where  $\overrightarrow{n_i}$ and $\overrightarrow{n_j}$ are term vectors for the nodes $n_i$ and $n_j$ respectively. The weight of each term in $\overrightarrow{n_i}$ is calculated as $tf*isf$. Here $tf$ is \textit{term frequency} and $isf$ is \textit{inverse sentential frequency}. $term~frequency$ is defined as the number of times a term occurs in a sentence. \textit{inverse sentential frequency} is defined as \begin{math} log(\frac{N}{n_{t}+1}) \end{math}, where $N$ is total number of sentences in the document and $n_t$ is number of sentences in which the term is present. 

In this section we propose an approach to embed the summary into the new document. Algorithm \ref{Algo:UpdateDocument} sketches the details of the embedding of the current summary into the new document. 
\begin{algorithm}[]
\caption{To embed summary into the document}
\begin{algorithmic}[1]\label{Algo:UpdateDocument}
\STATE \textbf{Input}: CurrentSummary and NewDocument   
\STATE \textbf{Output}: Document with summary embedded into it  
\IF {size(CurrentSummary) $\geq$ size(NewDocument)}
	\STATE Swap CurrentSummary and NewDocument \label{6}
\ENDIF
\STATE Let $d_1,d_2.....d_y$ be the nodes in document\\ \COMMENT{//No. of nodes in document = "y"}
\STATE Let $s_1,s_2.....s_x$ be the nodes in summary \\ \COMMENT{//No. of nodes in the summary = "x"}
\STATE EmbeddedDocument = NewDocument \label{1}
%\STATE DocumentEmbedded = null
%\STATE tempNode = null
\STATE Insert the last sentence of the summary into the EmbeddedDocument(all the nodes in the  EmbeddedDocument are considered for insertion) using the strategy explained in Section \ref{InsertionStrategy} \label{2}
\STATE Insert the first sentence of the summary into the EmbeddedDocument(only the nodes above the $s_x$ in the EmbeddedDocument are considered for insertion)  using the strategy explained in Section \ref{InsertionStrategy} \label{3}
\WHILE {All the nodes of summary are not embedded into the EmbeddedDocument (starting from $s_2$)}
	\STATE \COMMENT{// Consider the insertion in the summary order} \label{4}
	%\IF{NextSummary $\neq$ null}
	\STATE Insert the summary node $s_i$ into the EmbeddedDocument(only the nodes between $s_{i-1}$ and $s_x$ in EmbeddedDocument are considered for insertion) using the strategy explained in Section \ref{InsertionStrategy} \label{5}
	%\ENDIF
\ENDWHILE	
\STATE Return EmbeddedDocument  
\end{algorithmic}
\end{algorithm}
The Algorithm \ref{Algo:UpdateDocument} gives the method of embedding the sentences from summary into the document. Line 3 is very crucial, here the $size(D)$ gives the number of sentences in $D$. Idea is that if the size of the summary is less than the new document's size then the summary will be embedded into the new document otherwise the new document will be embedded into the summary. 
\subsection{Insertion Strategy}
\label{InsertionStrategy}
This section gives the detailed explanation of insertion strategy. A node $s$ in the summary is placed in the document appropriately. 
The steps to be followed are given below:
\begin{itemize}
\item Similarity (calculated using Equation \ref{EdgeScore}) of $s$ is calculated with the nodes (the nodes that are specified in Algorithm \ref{Algo:UpdateDocument}) in the document. 
\item Let $y$ be a node in the document which has maximum similarity with the node in the summary. 
\item Let $x$ and $z$ be the preceding and following nodes of $y$ respectively.
\item Calculate the similarity of $s$ with $x$ and $z$.
\item $s$ is placed in between $x$ and $y$ if $s$ has greater similarity value with $x$ than $z$, otherwise $s$ will be placed in between $y$ and $z$.
\end{itemize}        
%The above strategy is applied even when $x$ and $z$ are null. 
\subsection{Handling Exceptions}
When the similarity value of node $s_i$ is zero with every node of the EmbeddedDocument then the node is inserted immediately after $s_{i-1}$ in EmbeddedDocument. Here node $s_i$ is the node that is following node $s_{i-1}$ in the summary. If $s_{i-1}$ is not present then the node is placed immediately before $s_{i+1}$ in the EmbeddedDocument (in this case, $s_{i+1}$ is inserted before inserting $s_i$). The former process is recursive in nature. Even after calling recursively if the nodes in the summary are not embedded then the summary will be appended to the document.
 
This exception handling module will be used rarely by the system. We assume that the new document which arrived is related to the topic and therefore it is unlikely that the sentences in summary will have similarity value of zero with the sentences in the new document. Even otherwise the strategy holds good i.e., if the new document is an outlier(document that does not contain any information related to the query) then none of the sentences will be selected from the new document and the sentences of old summary alone will be selected.         
\section{Update Summary Generation}
\label{Algorithm}
%querying the document
In this section, summary generation on the embedded document is discussed. Here the score of the node is calculated based on the query posed by the user i.e., the node gets score based on its relevance to the query. 
\subsection{Node Score}
\label{Sec:NodeScore}
Node score calculation is based on the Equation \ref{NodeScore}. 
\[f(n,q_i) = \left\{ 
\begin{array}{l l}
  1/t & \quad \mbox{if $q_i$ is present in $n$}\\
  0 & \quad \mbox{if $q_i$ is not present in $n$}\\
\end{array} \right. \]
Here $t$ is the number of query terms in the given query, $n$ is the sentence and $q_i$ is the query term. If the query term is present in the sentence then a non-zero value is assigned otherwise zero is assigned.
\begin{center}
\begin{equation}
\label{NodeScore}
%\begin{split}
       w_{q_i}(s) = d*f(s,q_i)  +  \frac{(1-d)}{a}\sum_{v\in adj(s)} sim(s,v)*f(v,q_i) 
%\end{split}       
\end{equation}
\end{center}
Here $a$ is the number of sentences adjacent to $s$ that have the query term and have non-zero similarity with $s$. $d$ is the bias factor. In Equation \ref{NodeScore}, the first part captures the importance of the sentence with respect to the query term and the second part captures the type of neighbours (adjacent sentences). Two sentences are said to be adjacent if the similarity value between them is above a threshold(=0.001). \textit{The score of a node is the summation of Equation \ref{NodeScore} over all the query terms.} Unlike the node score equation in \cite{QueSTS}, the Equation \ref{NodeScore} is not iterative. Also, this equation considers only immediate neighbours while assigning node scores. This makes the system efficient. 
  
\subsection{Summary Generation}
Node scores are calculated for all the nodes and summary generation is explained in this section.
\begin{algorithm}[]
\caption{Generating Summary}
\begin{algorithmic}[1]\label{Algo:SummaryGeneration}
\STATE \textbf{Input}: EmbeddedDocument   
\STATE \textbf{Output}: Summary
\STATE SUMMARY = null
\STATE COUNT = null
\STATE Select the highest scored node in EmbeddedDocument into the SUMMARY\label{11}
\WHILE{All the query terms are not included into the SUMMARY AND (COUNT != SummarySize)}
\STATE Recalculate  the scores of the nodes using Equation \ref{MMR1}\label{12}\COMMENT{//This calculation is only for temporary purpose. At the beginning of each iteration the nodes are assigned their original node scores}
\STATE Select a node into SUMMARY that maximizes number of query terms in the SUMMARY\COMMENT{//IF more than one such node is present then select the node that has maximum score}\label{13}
\STATE COUNT++ \label{14}
\ENDWHILE
\WHILE{COUNT != SummarySize}
\STATE Select the next highest scored node from EmbeddedDocument using Equation \ref{MMR2}\label{15}
\STATE Add the highest scored node to SUMMARY 
\STATE COUNT++
\STATE Calculate temporary node scores using Equation \ref{MMR1}
\ENDWHILE
\STATE Return SUMMARY 
\end{algorithmic}
\end{algorithm} 
In Algorithm \ref{Algo:SummaryGeneration}, it is assumed that $SummarySize$ (number of sentences that user wants as a summary) is not greater than the number of sentences in the EmbeddedDocument. From Lines 7 to 9, the completeness of the summary is achieved. A summary is complete if all the query terms are present in it. Then the nodes are added from the remaining pool as given in Lines 12 to 15. In Line 7, Equation \ref{MMR1} is used to recalculate the node scores and in Line 12 the maximum scored node is selected using the Equation \ref{MMR2}. Note that here, scores are assigned to nodes temporarily using Equation \ref{MMR1} and Equation \ref{MMR2} is used to select the highest scored node into summary. After the selection, the node scores are reverted to their original scores(as calculated in Section \ref{Sec:NodeScore}). 

\begin{equation}
\label{MMR1} 
tempW_{Q}(n_i) = \kappa \lambda \sum_{1\leq k \leq t}w_{q_k}({n_{i}})-(1-\lambda)\underset{j}{Max}\lbrace sim(n_{i},s_{j})\rbrace
\end{equation}  

\begin{equation}
\label{MMR2} 
\underset{i}{Max}\lbrace tempW_{Q}(n_i) \rbrace 
\end{equation}       

Here $n_i$ and $s_j$ represents document and summary nodes respectively. $tempW_{Q}(n_i)$ is the temporary node score of $n_i$. Equation \ref{MMR1} is inspired from \cite{MMR}. The sentences in the summary generated using Algorithm \ref{Algo:SummaryGeneration} are rearranged in the document order. This summary is complete, coherent and also non-redundant. The value of $\lambda$ is taken from \cite{MMR}. $\kappa$ is used as a scaling factor and it is fixed empirically.  
\subsection{Discussion}
\subsubsection*{Complete Summary}
While selecting sentences into summary, the sentences which will cover maximum uncovered query terms are given highest preference. The generation of summary is carried out by adding one sentence followed by another. Fist sentence which is included into summary will be the highest scored sentence. The sentences selected after that are targeted towards maximizing the number of query terms coverage. If more than one sentence is contributing the same number of query terms then the highest scored sentence among them will be selected to be included into the summary. This process is repeated till all the query terms are included into the summary. 
\subsubsection*{Coherent Summary} 
The sentences selected into the summary are arranged in the EmbeddedDocument order. The insertion strategy discussed in Section \ref{Model} ensures that the EmbeddedDocument is coherent i.e., the sentences in the EmbeddedDocument are well connected and there is a logical flow within sentences. Therefore updated summary is coherent.  
\subsubsection*{Quality Summary}
After achieving the task of complete summary, the nodes that are included are purely based on two criteria: First one is the node's importance w.r.t the query and second is its contribution to the summary. Contribution is the amount of new information it is adding to the summary. In other words it is non-redundancy. So, Equation \ref{MMR2} is used to select the sentences which ensures the non-redundancy and thus the quality of the summary. Recall that before selecting the highest scored node, Equation \ref{MMR1} is used to calculate the temporary node scores. 
\subsubsection*{Efficiency}
In this system we embed the summary into new document in a coherent manner and then the summary is generated by extracting sentences from the embedded document. The complexity of the system is $O((S_i+D_j)^2)$, $S_i$ and $D_j$ are number of sentences in current summary and new document respectively. The complexity of a multi-document summarizer is $O( (\sum D_j)^2)$.             
\section{Experimental Setup}
\label{ExperimentalSetup}
Evaluating the proposed system is a difficult task. Update summary generation is evaluated on DUC 2006\footnote{http://duc.nist.gov} corpus. DUC has 50 topic clusters and each topic is described in 25 documents. Initial summary is generated using the MEAD \cite{MEAD} system for the query and the document cluster provided by DUC. This summary is generated on the first 15 of the 25 documents. The summary generated will be the input for the update summary generation task. The $16^{th}$ document will be the new document into which the summary is to be embedded. The summary is generated for the given query on the embedded document and this generated summary will be embedded into $17^{th}$ document. The process is repeated till the summary on the last embedded document($25^{th}$) is generated. 

\begin{figure}[h]
\begin{center}
\centering
\includegraphics[scale = .45]{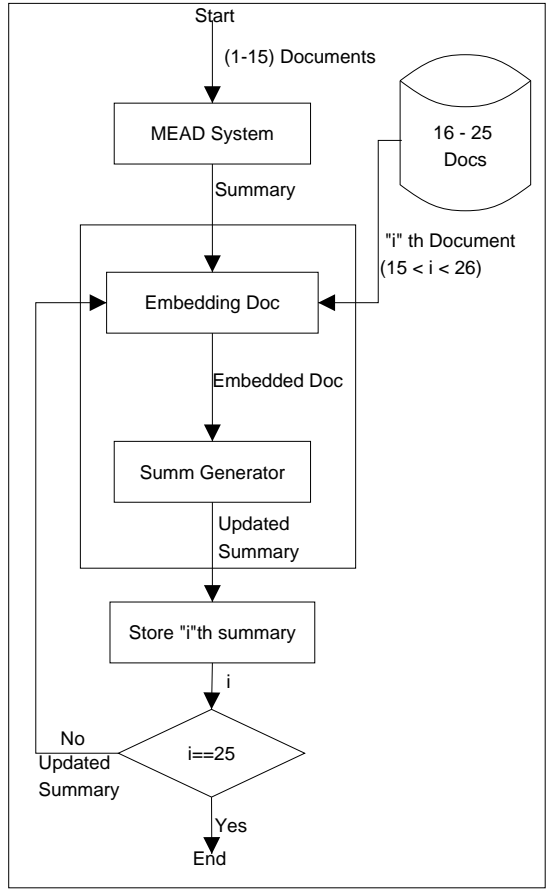}
\end{center}
\caption{A block diagram of experimental setup}
\label{ExperimentalSetupfig}
\end{figure}

The block diagram for the experimental setup is shown in Figure \ref{ExperimentalSetupfig}. MEAD \cite{MEAD} follows centroid based approach to generate summaries. It deals with both single and multi-document summarization. In our setup we use MEAD's multi-document summarization approach. MEAD computes a score for each sentence from the given cluster of related documents by considering a linear combination of several features. We have used centroid score, position and cosine similarity with query as features with 1,1,10 as their weights respectively. MMR(Maximum Marginal Relevance) re-ranker is used for redundancy removal with a similarity threshold of 0.6. 
The updated summaries so formed are all stored and evaluated against the model summaries given by DUC. In DUC, the model summaries are of fixed length i.e., 250 words. So, all the generated summaries are truncated to 250 words.   
\subsection{Discussion on Baseline Summaries}
This problem is first posed by us and therefore there is no other system available to be compared with the performance of our system. As this is an update summary generation task there is no meaningful baseline system that can be compared with our system. The following alternatives were thought of for a baseline system: 1) Generate a baseline summary using MEAD with all the 25 documents as input. As our system generates the summary by considering only the current summary and new document, this is not a fair comparison. 2) If baseline summary for $i^{th}$ document inclusion is available then baseline summary for $i+1^{th}$ document inclusion can be calculated using MMR approach. But the former approach requires the presence of all the $i+1$ documents to generate a baseline summary. So, it also will not be appropriate baseline. 

So, we give the ROUGE results generated by the best performing system of DUC 2006(System-24), these values are for summaries generated by considering all the 25 documents. But USUM's ROUGE values are \textit{not} obtained by considering all the 25 documents. So, the values of the best system of DUC 2006 would naturally be better than our systems values.
\section{Experimental Results}
\label{ExperimentalResults}

The ten updated summaries for each cluster are evaluated according to DUC 2006 specifications. DUC uses ROUGE measures to evaluate the quality of the summary generated by comparing with the model summaries. Recall is calculated for the generated summaries w.r.t this model summaries. ROUGE\cite{ACL2004ChinYewLin} stands for Recall-Oriented Understudy for Gisting Evaluation. ROUGE measures the quality of a summary by comparing it to the summaries created by
volunteers. ROUGE-N is n-gram recalls between system generated summaries and the summaries generated by the volunteers(models). ROUGE-N is calculated based on the Equation  \ref{ROUGEEquation}
\begin{equation}
\label{ROUGEEquation}
%\begin{split}
ROUGE\!-\!N = \frac{\underset{s \in model~summaries}\sum{}~~\underset{gram_n \in s }\sum{count_{match}(gram_n)}}{\underset{s \in model~summaries}\sum{}~~\underset{gram_n \in s }\sum{count(gram_n)}} \\
%\end{split}
\end{equation} 
Here $n$ is the length of n-gram. $gram_n$ stands for n-gram. $Count_{match}(gram_n)$ is the maximum number of n-grams co-occurring in both the generated summary and in the reference summaries. ROUGE-1 and ROUGE-2 are the recall measures of unigrams and bi-grams respectively. ROUGE-W is the weighted longest common subsequences matching. In longest common subsequence matching, the distance between the words is not considered as an important issue but in weighted longest common subsequence matching, weight is given to the distance between the words. ROUGE-SU4 is the recall measure which computes the skip bi-grams with skip distance four and uni-grams are also considered while computing this measure. 

In Table \ref{tab:4} the ROUGE values for updated summaries generated on DUC 2006 are given. The values in the table are averaged values over 50 clusters. Updated summary 1 is the summary obtained by updating the summary generated on first 15 documents with the sixteenth document. Updated summary 2 is the summary obtained by updating the updated summary 1 with the seventeenth document. We empirically found that the ROUGE values are better for $\kappa$ value of 20. We also give the ROUGE values for the system-24(best performing system) of DUC 2006 in Table \ref{BestSystem}. The values in Table \ref{BestSystem} are for the summaries generated by considering all the 25 documents of the cluster. So, the ROUGE values of Table \ref{BestSystem} will be better than the ROUGE values of our systems. But the ROUGE values of our system are very close to the ROUGE values of the system-24. This indicates that our system is performing well.    

The proposed system is implemented on the system with the following configuration: 256MB main memory, 1.7 GHz Intel Pentium processor and the operating system is FC3. The system is implemented in Java. The time taken to compute the update summaries on 50 clusters is 56 minutes. So, it is slightly greater than 1 minute per cluster. On average it is less than 7 seconds per update summary(there are 10 update summaries per cluster).    

\begin{table*}
\centering
\caption{ROUGE Values on DUC 2006 with $\kappa$ value 20}
\begin{tabular}{|c|c|c|c|c|}  \hline  
Updated Summary&ROUGE-1&ROUGE-2&ROUGE-W&ROUGE-SU4\\ \hline
1&0.38980 & 0.08179 & 0.09429& 0.13757\\ \hline
2&0.38660  &0.07905  &0.09321 &0.13552 \\ \hline
3& 0.38919& 0.08196& 0.09418 & 0.13786\\ \hline
4& 0.38871& 0.08239& 0.09351& 0.13713\\ \hline
5& 0.38457& 0.08024& 0.09274& 0.13472\\ \hline
6& 0.38467& 0.08060& 0.09297& 0.13490\\ \hline
7& 0.38547& 0.08058& 0.09339&0.13518 \\ \hline
8& 0.38282& 0.08004& 0.09245& 0.13389\\ \hline
9& 0.38358& 0.07955& 0.09281& 0.13390\\ \hline
10& 0.38432& 0.08031&0.09282 & 0.13419 \\ 
\hline\end{tabular}
\label{tab:4}
\end{table*}
\begin{table*}
\centering
\caption{ROUGE Values of System24 on DUC 2006}
\begin{tabular}{|c|c|c|c|c|} \hline
Updated Summary&ROUGE-1&ROUGE-2&ROUGE-W&ROUGE-SU4\\ \hline
System24&0.41108 &0.09558 &0.11068 &0.15529 \\ 
\hline\end{tabular}
\label{BestSystem}
\end{table*}
\section{Conclusions} 
\label{Conclusions}
In this paper, the current summary is cleverly embedded into the new document in a meaningful and coherent way. A query specific summary is generated on the embedded document. The sentences which are extracted from the document form a \textit{complete} summary. The algorithm proposed will not select sentences which have redundant information. All the sentences are arranged in the embedded document order to maintain the coherence and flow in the summary. The system is efficient and the quality of the update summary is satisfactory. The results are highly encouraging. USUM gives efficient solution for update summary generation which is a challenging and useful task.      
% \bibliographystyle{splncs}
%  \bibliography{references}

\end{document}